\documentclass[printer]{aa} 
\usepackage{graphicx}
\usepackage{epstopdf}
\usepackage{txfonts}
\def\d#1{{\mbox{d}{#1}}}		

\begin{document}
   \title{Spatial distribution of interstellar dust in the Sun vicinity}

   \subtitle{Comparison with neutral sodium-bearing gas}

   \author{J-L. Vergely \inst{1},
          B. Valette \inst{2},
          R. Lallement \inst{3},
           \and
           S. Raimond \inst{3},
         }

   \institute{  1 - ACRI-ST, 260 route du Pin Montard, Sophia Antipolis, France\\
                \email{jeanluc.vergely@aerov.jussieu.fr}\\
                2 - LGIT-Savoie, Universit\'e de Savoie, 73376 Le-Bourget-du-Lac,
                France \\
                3 - LATMOS, 91371 Verri\`eres-Le-Buisson, France\\
             }

   \date{Received Dec, 23, 2009; revised Feb, 24, 2010 }


  \abstract
   {}
   {3D tomography of the interstellar dust and gas may be useful in many respects, from the physical and chemical evolution of the ISM itself to foreground decontamination of the CMB, or various studies of the environments of specific objects. However, while spectral data cubes of the galactic emission become increasingly precise, the information on the distance to the emitting regions has not progressed as well and relies essentially on the galactic rotation curve. Our goal here is to bring more precise information on the distance to nearby interstellar dust and gas clouds within 250 pc.}
   {We apply the best available calibration methods to a carefully screened set of stellar Str\"omgren photometry data for targets possessing a Hipparcos parallax and spectral type classification. We combine the derived interstellar extinctions and the parallax distances for about 6,000 stars to build a 3D tomography of the local dust. We use an inversion method based on a regularized Bayesian approach and a least squares criterion, optimized for this specific data set. We apply the same inversion technique to a totally independent set of neutral sodium absorption data available for about 1700 target stars.}
   {We obtain 3D maps of the opacity and the distance to the main dust-bearing clouds with 250 pc and identify in those maps well-known dark clouds and high galactic more diffuse entities. \\
   We calculate the integrated extinction between the Sun and the cube boundary and compare with the total galactic extinction derived from infrared 2D maps. The two quantities reach similar values at high latitudes, as expected if the local dust content is satisfyingly reproduced and the dust is closer than 250 pc. Those maps show a larger high latitude dust opacity in the North compared to the South, reinforcing earlier evidences. Interestingly the gas maps do not show the same asymmetry, suggesting a polar asymmetry of the dust to gas ratio at small distances.\\
   We compare the opacity distribution with the 3D distribution of interstellar neutral sodium resulting from the inversion of sodium columns. We discuss the similarities and discrepancies and the influence of  data set limitations. Finally we discuss the potential improvements of those 3D maps.
}
   {}


   \maketitle
%

\section{Introduction}

The 3D representation of the local ISM is an important challenge
in various perspectives. In the most direct way it allows a better modelling of the ISM. The ability to localise masses of gas and dust, combined with information on the ionisation state, the temperature and the kinematics helps to understand the chemical and physical evolution of the different gas components in relation with their history and the surrounding ionizing radiation from hot stars, as well as the dust-gas relationships and the dust processing (Cox, 2005; Draine, 2003). \\
In a more indirect way, the 3D ISM structure becomes increasingly useful  as an ingredient to the foreground contamination of the CMB because the local distribution of the gas influences the interstellar radiation field, which in turn influences the dust temperature and infrared-millimetric emission. The 3D distribution of the local ISM is also governing a significant fraction of the diffuse gamma-ray background and is one of the ingredients of the models developed in the context of the Fermi satellite.\\
In a very different way, the knowledge of the 3D properties of the local ISM may be useful in various other studies since it provides the environmental context of astrophysical objects, allows a correction of the extinction according to the distance and the direction, allows to disentangle spectral lines in absorption related to the object itself or to the foreground, or provides a tool for the estimate of the propagation of energetic particles. The list of topics is too long to be detailed here. \\

Recent work on the 3D global structure and kinematics of the local ISM  or on specific nearby ISM structures has been done
using absorption lines in the UV (e.g. Lallement et al, 1995, Redfield and Linsky, 2008, Welsh and Lallement, 2005) and the strong NaI and CaII interstellar  in the visible (e.g. Genova and Beckman, 2003, Snow et al, 2008). A number of works have combined emission data and neutral sodium absorption to attribute a distance to specific features seen in radio or IR (e.g. Lilienthal et al, 1992, Corradi et al, 2004, Meyer et al, 2006), making use essentially of Hipparcos parallaxes (\cite{perryman}). Neutral species like NaI trace low temperature
regions, i.e. dense atomic gas clouds and molecular clouds, while ionised calcium is a tracer of both dense regions (at the exception of the fully neutral cores) and warm diffuse gas, partially ionised (see Welty et al, 1996 for more details). Note that the known existence of different typical sizes linked to the different types of gas, from parsecs for molecular clouds to tens of parsecs for atomic and diffuse clouds, has influenced our choice of correlation lengths for the inversion described in section 4.
In parallel, Str\"omgren photometry has been used  in combination with parallaxes for distance estimates of a number of features (e.g. Reis et Corradi , 2008, Knude and H\"og 1998 , 1999)

 It is known for a long time from both absorption data and diffuse soft X-ray background detection that the Sun is embedded in a relatively empty cavity, the Local Bubble (LB), mostly devoid of dense clouds (Frisch and York, 1983, Sfeir et al, 1999). Within this local cavity reside essentially tenuous diffuse clouds such as
the clouds that make part of the so-called Local Fluff within 20 pc from the Sun. The interpretation of the soft X-ray background data suggests that the cavity is filled by one million K gas  (Snowden et al, 1990). The cavity is delimited by dense clouds whose distances range from 40 to 180 parsecs and neighboring cavities are distributed all around.
Such a "cellular" structure of hot bubbles separated by high density and relatively compact regions, including
molecular dust clouds is totally expected from the permanent ISM recycling that shapes it: stellar winds and supernova explosions inflate hot "bubbles" within the dense gas and maintain these high contrasts (Cox and Reynolds, 1987, De Avillez and Breitschwerdt, 2004).
There is very probably a strong  link between the strong winds having inflated the cavity and the surrounding so-called Gould belt, an expanding elliptical ring of star-forming regions (see Perrot and Grenier, 2003).
Several questions remain however about the local ISM and Local Bubble. The embedded clouds are ionized by the early-type stars, but their measured ionisation state, especially the ionization of helium requires an additional harder radiation supposed to be generated within semi-hot conductive interfaces between clouds and hot gas (e.g. Slavin 1989). The distribution of the corresponding semi-hot or hot gas ions however is not in fully satisfying agreement with this scenario (Welsh and Lallement, 2005, Savage and Lehner, 2006, Knauth et al, 2003).  On the other hand the more recent discovery of solar wind charge-transfer X-ray emission (Cox, 1998, Cravens, 2000) has also complicated the picture since this diffuse emission, which mimics one million K thermal emission, is as intense as the observed background in a number of directions  (e.g. Koutroumpa et al, 2009). Decreasing the hot gas pressure significantly would have the positive effect of suppressing the strong difference from the pressure measured within the diffuse clouds and the local cloud (Jenkins, 2009).\\
From 3D gas and dust mapping one can expect significant progresses on these topics,  the "Local Bubble" wall location and the X-ray emission source regions, some hints about the local bubble formation and the link between the local
bubble and surrounding bubbles and more generally a better understanding of the multi-phase structure and the interaction between the different phases of the ISM. One can also expect the determination of the regions of star formation and estimates of the cloud masses. Last but not least, it should help to shed light on the surprisingly high deuterium abundance in the local ISM (Linsky et al, 2006), now largely recognised to be related in a very large extent to dust evaporation from grains (Draine, 2004). The unambiguous correlation between refractory metals and deuterium abundances in the gas (Lallement et al, 2008) as well as the peculiar distribution of the deuterium enhancements with respect to the Gould belt (Lallement, 2009) suggest a scenario for the formation of the belt with a large role of dust and dust-gas processes. \\

The first computed tomography of the local ISM from \cite{vergely1} used neutral sodium absorption data combined with Hipparcos stellar parallaxes (Perryman et al, 1997) and a tomographic method based on the work of Tarantola \& Valette (1982a), specifically applied for the first time to extinction by the ISM  by \cite{chen}. This type of tomography is a new manner to represent the ISM, considering it as a continuous
medium without a priori cloud discretization and considering the clouds and the voids as fluctuations around a median density, the "prior" value.
\cite{lall03} presented  additional sodium observations especially dedicated to the Local Bubble boundary determination, recorded during several observing programs, and made use of about 1,000 carefully screened data for a column density mapping and a tomography based on the same algorithm as in \cite{vergely1}. An updated study using about 1700 targets as well as the first tomography using ionized calcium are presented in the recent study by Welsh et al (2009). Increasing the number of targets has allowed to reveal neighboring cavities and to gain in precision. \\
Using a similar approach, we present here the 3-D density
distribution of the extinction in the vicinity of the Sun,
using color excesses from Str\"omgren photometry and Hipparcos
distances. Because extinction traces low temperature areas as
does neutral sodium, a comparison of the derived extinction maps with the
results of Welsh et al. 2009 is performed.

Section 2 describes in detail the determination of the extinction and the associated uncertainties. Section 3 describes the basis of the inversion method and the algorithm used in our study as well as the underlying assumptions. Section 4.1 compares the total extinction towards the target stars with the total extinction deduced from IR data by Schlegel et al (1998), and compares integrated extinctions and NaI columns. Section 4 shows the results of the inversion in several planes and the comparison with the inversion of the neutral sodium absorption. Section 5 presents the conclusions and discusses shortcomings and potential improvements of the inversion and the database.

  \begin{figure*}
   \centering
   \includegraphics[width=\textwidth]{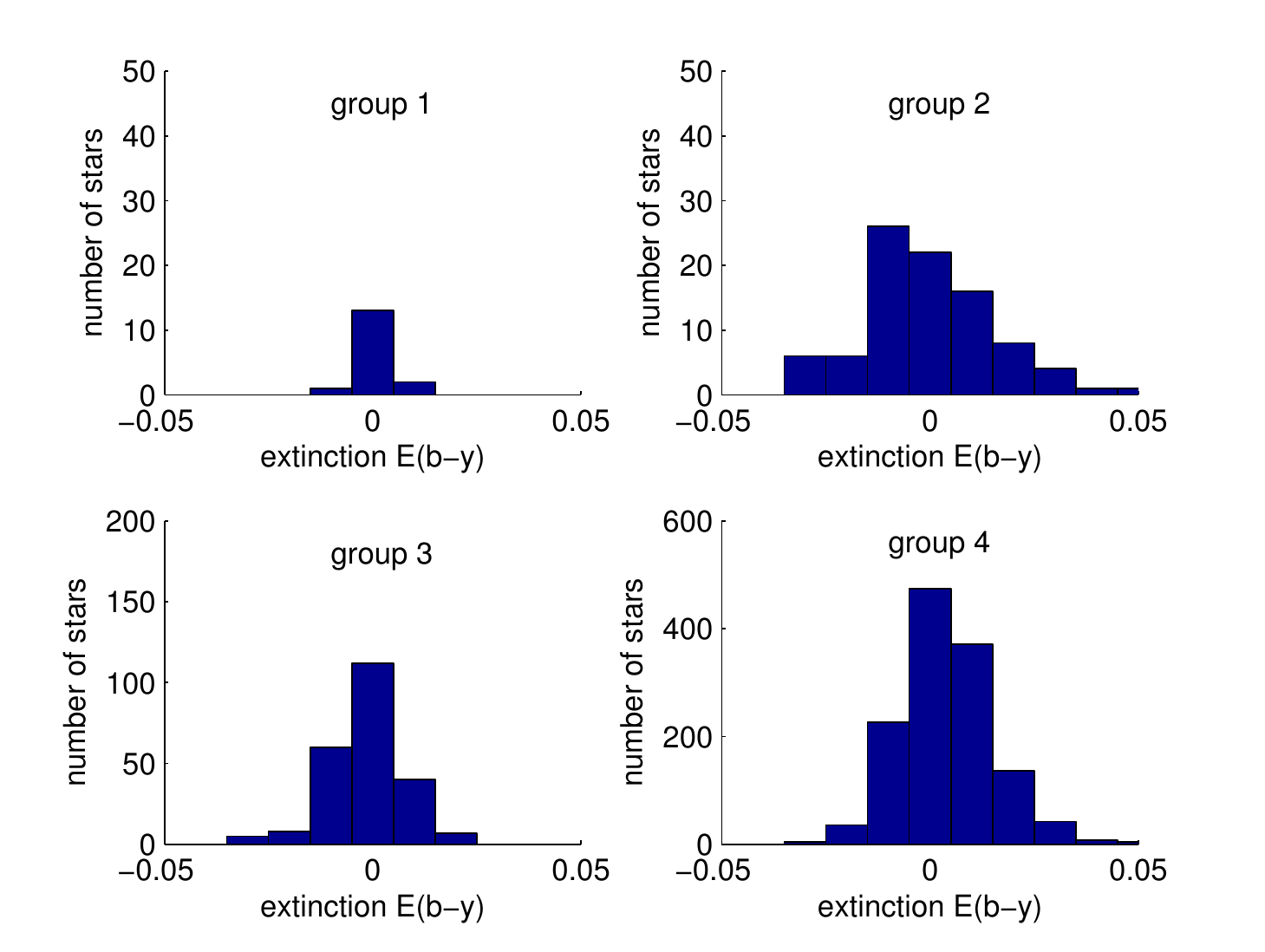}
      \caption{Histogram extinction for stars in the 70 pc sphere}.
         \label{FigHistoExt}
   \end{figure*}

\section{Photometric data and extraction of extinction}

\subsection{Extinction representation}
In the standard model of stellar light absorption by intervening dust clouds, the intensity decreases exponentially along the line-of-sight as a function of the opacity :
  \begin{equation}
        I(\lambda)=I_0(\lambda)\exp\left( -\tau_{\lambda} \right)
   \end{equation}
$I(\lambda)$, $I_0(\lambda)$
are the absorbed and unabsorbed intensities and $\tau_{\lambda}$ is the optical depth, or opacity. $\tau_{\lambda}$ depends on the wavelength $\lambda$, the absorption coefficient
   $k_{\lambda}$ and the density of the dust $\rho_{dust}$, and is the integral along the line-of-sight (LOS) up to the distance $R_i$ of the star:

   \begin{equation}
        \tau_{\lambda}=\int_0^{R_i} k_{\lambda}(r) \rho_{dust}(r) \d r
  \end{equation}

The observed flux
decreases with the dust absorption and with the distance. To quantify the dust absorption solely and remove a dependence on the intrinsic stellar flux and the distance one uses the color excess
$E_i(\lambda_1, \lambda_2)$  that is the logarithm of
the intensity ratio between two different bands $\lambda_1$ and
$\lambda_2$ :
 \begin{equation}
        E_i(\lambda_1, \lambda_2) = \mbox{C} \int_0^{R_i} \left( k_{\lambda_1}(r) - k_{\lambda_2}(r) \right) \rho_{dust}(r) \d r
  \end{equation}
In the following sections we adopt the notation :
  \begin{equation}
        E_i := E_i(\lambda_1, \lambda_2) \mbox{  and  } k := k_{\lambda_1} - k_{\lambda_2}
  \end{equation}

Note that $k_{\lambda}(r)$ depends on the composition of the
dust which can vary along the line of sight.
However our purpose here is to estimate the opacity and not to
extract the actual dust density by using an extinction coefficient.
Thus, the variable of interest is the differential opacity, a combination of the dust properties (dust size distribution and extinction coefficient)
and the dust density:

\begin{equation}
        \rho_{op}(r) := \mbox{C} \left(k_{\lambda_1}(r) - k_{\lambda_2}(r) \right) \rho_{dust}(r)
   \end{equation}

 In this case
 \begin{equation}
  E_i(\lambda_1, \lambda_2) =  \int_0^{R_i} \rho_{op}(r) \d r
  \label{equa_fond}
  \end{equation}
   In this paper we use the differential opacity between the Str\"omgren photometry B and y bands:

\begin{equation}
        \rho_{op}(r) := \mbox{C} \left(k_{\bf{b}}(r) - k_{y}(r) \right) \rho_{dust}(r)
   \end{equation}
The quantity $\rho_{op}$ is expressed in color excess per pc.

The tomography of the extinction assumes a continuous distribution of the opacity in 3D and requires a 3D formulation of
Eqn. \ref{equa_fond}. For each line-of-sight $i$ among the $n$ measured opacities constituting the data set, the $i$th constraint is expressed in the following way over the whole space V:

   \begin{equation}
    \label{equa_fond2}
        E_i= \mbox{C}\int_V K_i(x) k(x)\rho_{dust}(x) \d V(x) = \int_V K_i(x) \rho_{op}(x) \d V(x)
   \end{equation}

  where x is the position vector in galactic coordinates : $x =(r,l,b)$ with $\d V(x) =
  r^2 cos(b)\d r \d b \d l$, and where the singular kernel $K_i( x)$ is explicitly
   given by :

   \begin{equation}
        K_i(r,l,b)=\Theta_i(r)\delta(b-b_i)\delta(l-l_i)
   \end{equation}
   in denoting by $\Theta_i(r)$ the function equal to $1/r^2cos(b_i)$ inside the interval $[0,R_i]$
   and null outside, and by $\delta$ the Diracs' distribution.

The density of IS matter decreases
strongly with the distance to the galactic plane. A convenient
description of this behaviour is to suppose {\it a priori} an
exponential decrease with height above the galactic plane (Chen
et al., 1998) :
  \begin{equation}
  \label{rhomean}
\rho_{ref}(r,b)=\rho_{0}\exp\left(-\frac{|r\sin(b)|}{h_0}\right)
  \end{equation}
We assigned a value of 200 pc to $h_0$ for both the opacity and NaI.
Note that for dust and gas 200 pc is somewhat larger than what is usually used
 (100 pc for the dust, see, i.e., Chen et al., 1998, and 160 pc for NaI as deduced from a fit to the column-densities
 of our NaI database), but we have done it intentionally
 in order to avoid the disappearance of the tenuous high latitude structures along the inversion
when the "a priori" solution is too small.\\
In this study, we propose to find an extinction model close to
the reference model given in Eqn. \ref{rhomean} that explains
observed star extinctions.\\
Since the density is a positive quantity, it is convenient to make the following
change of variable :
  \begin{equation}
\rho(x)=\rho_{ref}\exp\left(\alpha(x) \right)
  \end{equation}
where the log opacity $\alpha$ is the new unknown parameter, which
is assumed to follow a gaussian centered law, and which
traces the difference in opacity from the reference model. In
the following numerical applications, we have taken an average
E(b-y) color excess $\rho_{0}=0.0004$ per $\mbox{pc}$. This choice, which corresponds to about twice the average measured value, is motivated by the following reasons: i) we aim at a robust result  on the very low opacity within the Local Bubble. Obtaining a strong decrease  from the "a priori" model demonstrates the observational constraints on the cavity. ii) we aim at showing here that the inversion algorithm does not depend on the " a priori" opacity. iii) in areas with high dust concentration there may be an observational bias towards weaker opacities due to the limited sampling (lines of sight towards the targets do not cross the densest cores). Starting with larger opacities may partially compensate for this effect.\\

\subsection{Extinction determination from Str\"omgren photometry}

For the derivation of color excesses we use the Str\"omgren
color indices from the Hauck-Mermilliod Catalogue
(\cite{Hauck}). These authors have produced an extensive
compilation of photometric data, and in case of multiple
measurements  they have computed the averaged values, taking
into account the individual errors. The catalog lists 66,000
objects.  We have selected from the catalog those stars
possessing H$\beta$ determinations and Hipparcos parallaxes higher
than 3.34 mas (about 16,000 stars). We have also very carefully
screened the data using the Tycho catalog and removed all stars
suspected or observed to be part of a multiple system, all
variable stars and all stars suspected to be surrounded by a
shell. Using the Tycho spectral type classification all
luminosity class I and II have finally also been removed. The
remaining data set (about 6400 targets) has then been entered
in the calibration process.

Using calibration relations and spectral type classification,
color excesses $E(b-y)$ were derived for four groups of stars.
Table 1 gives for the spectral type and the range of $\beta$
index for the different calibration relations used in this
study. A detailed description of the calibrations is given in
\cite{vergely1}. Because the tomographic technic is very
sensitive to the presence of outliers, a specific filter has
been applied on color excesses from group 1. As mentioned by
Crawford (\cite{crawford}), target stars with high $c1$ values
(evolved stars with high luminosity) should be removed. This is
due to the presence of a turn off in the (c1,(b-y)) color
diagram which generates an ambiguity in the location of the
zero age main sequence. Indeed, a careful inspection of stars
with $c1 > 0.9$ shows clearly the
presence of underestimated $E(b-y)$ color excesses and we have removed the corresponding targets.
Note that the threshold recommended from Crawford  (\cite{crawford}) is 0.58 instead of 0.9, which may question our use of the class III stars. However 
only 5 stars (from 6,400) do not fulfil this limitation. We have checked that those stars, which are found to have a very small color excess, show no disagreement with the surrounding targets.

Assuming that the expected $E(b-y)$ color excess is negligible
in the 70 pc sphere, it is possible to determine, for each
calibration group, the intrinsic dispersion and the bias of the
observed color excesses. The figure \ref{FigHistoExt} shows
that only slight biases affect the different group. In order to
avoid a propagation of such a bias in the tomography output,
this bias has been corrected. The table 2 gives the bias and
the standard deviation of the color excess from star belonging
to the 70 pc sphere. The standard deviation comes from
intrinsic dispersion in color-magnitude diagram and photometry
error. It gives us the possibility to estimate the color excess
error to be applied during the inversion.\\
We have checked our results on the extinction based on the 1997
Hauck Mermilliod Str\"omgren data base by comparing them to
previous results by Philip and Egret (1980). These authors have
used similar calibration relations for group 1, 3 and 4 and the
earlier Hauck-Mermilliod (1980) catalogue. The cross
correlation between our sample and the Philip and Egret sample
allows to extract 2456 common stars. The standard deviation
between the differential opacities  E(b-y) from Philip and
Egret and those from the present  study reaches 0.009 mag and
the mean difference is -0.003 mag. Most of those differences
are due to differences in the Str\"omgren photometry colors
which are more recent in the present study and in differences
in calibration relations (for intermediate group, we use
Hilditch et al.(1983) calibration).

\begin{figure}
\centering
\includegraphics[width=9cm]{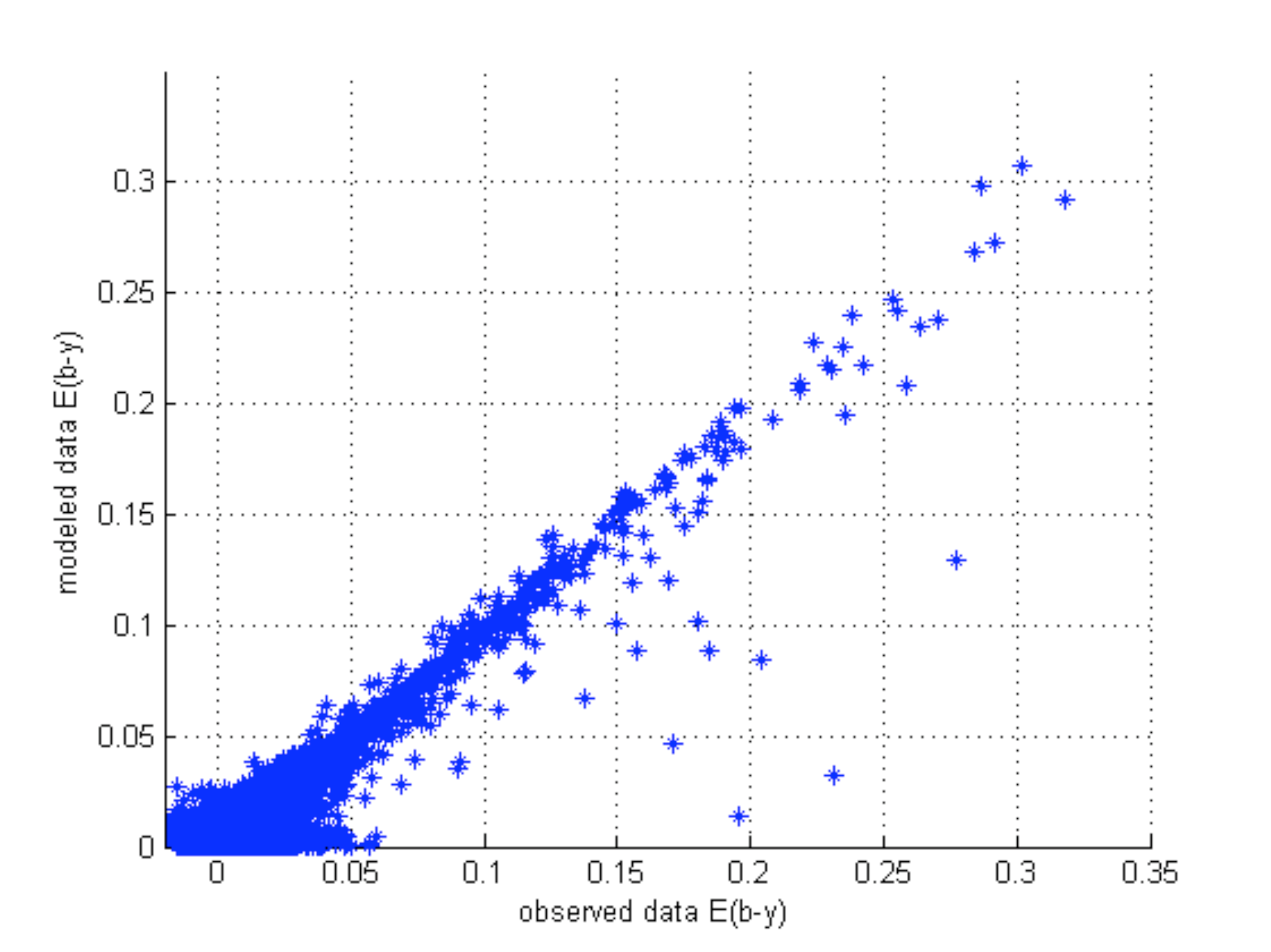}
  \caption{Fit between modeled data and observed one.}
     \label{fit}
\end{figure}

 \begin{table*}
      \caption[]{Distribution of the sample stars into 4 Str\"omgren groups (before parallaxe selection).}
         \label{}
    \begin{tabular}[h]{cccccc}
            \hline
group&spectral type&luminosity class&$\beta$&number of stars&calibration\\
           \hline
  1&    B0 to A0&  III to V&   $2.59<\beta<2.88$&   993&
Crawford (1978)\\
  2&    A0 to A3&  III to V&   $2.87<\beta<2.93$&    626&
Hilditch et al.(1983)\\
  3&    A3 to F0&  III to V&   $2.72<\beta<2.88$&   1517&
Crawford (1979)\\
  4&    F0 to G2&  III to V&   $2.56<\beta<2.72$&   3857&
Crawford (1975)\\
           \hline
 TOTAL&           &          &                 &   6993&
                 \\
           \hline
      \end{tabular}
   \end{table*}

\subsection{Target distances and distribution}
The distance to the stars affected by extinction in the solar
neighborhood is given by the Hipparcos parallaxes. In this
study, we use a new set of Hipparcos parallaxes consistently
derived by \cite{vanleeuwen}. This new Hipparcos
processing provides parallaxes which are slightly different but
of importance for our study is the resulting accuracy, claimed
to be a factor of 2.2 better than the initial one (from
\cite{perryman}). Because the distance accuracy reaches about
100 pc at 400 pc, we have kept stars closer than 300 pc, hence our limit on the parallax quoted above.\\
The distribution of retained targets is far from regularly
distributed in 3D. On the contrary, there are areas of high
concentration, and generally there is a strong drop of target
density beyond about 100 parsecs. The distribution of targets
is illustrated in the figures presented in section 5.

\subsection{Error estimates}

If observed data are biased or if their errors have been
underestimated, we cannot expect to obtain significant results
because the inversion technique tries to
reproduce an image of the initial data.
Data errors arise both from extinction and distance errors.
In our model, we do not consider these two errors as
independent but we propagate the distance error on that of the extinction.
The distance error $\sigma_d$ is related to the error of the
parallaxes, $\sigma_{\pi}$ in the following way :
$$
\vert\frac{\sigma_d}{d}\vert=\vert\frac{\sigma_{\pi}}{\pi}\vert
$$
The extinction error $\sigma_{{\rm E}d}$ resulting from the
distance error is approximately :
$$
\sigma_{{\rm E}d}={\rm E(b-y)}\, \frac{\sigma_d}{d}
$$
if we assume that the opacity is constant along the line of
sight. \\
So, if we consider a residual error of $\sigma_{cal}$ for the
extinction calibration, the total error, $\sigma_{{\rm E}T}$,
will be in the gaussian case:
$$
\sigma_{{\rm E}T}=\sqrt{\sigma_{{\rm E}d}^2+\sigma_{cal}^2}
$$
Table \ref{tabErrBias} shows the calibration error deduced from
the standard deviation of the stars belonging to the 70 pc
sphere.\\

\begin{table*}
      \caption[]{Estimation of errors and biases on extinction according to the group of calibration, for stars in the 70pc sphere}
         \label{tabErrBias}
    \begin{tabular}[h]{ccc}
            \hline
group&Error&Bias\\
           \hline
  1&    0.0044&  -0.0024 \\
  2&    0.0145&  -0.0011\\
  3&    0.0088&  -0.0009\\
  4&    0.0117 &  0.0040 \\
           \hline
      \end{tabular}
   \end{table*}

 \begin{figure}
\centering
\includegraphics[width=9cm]{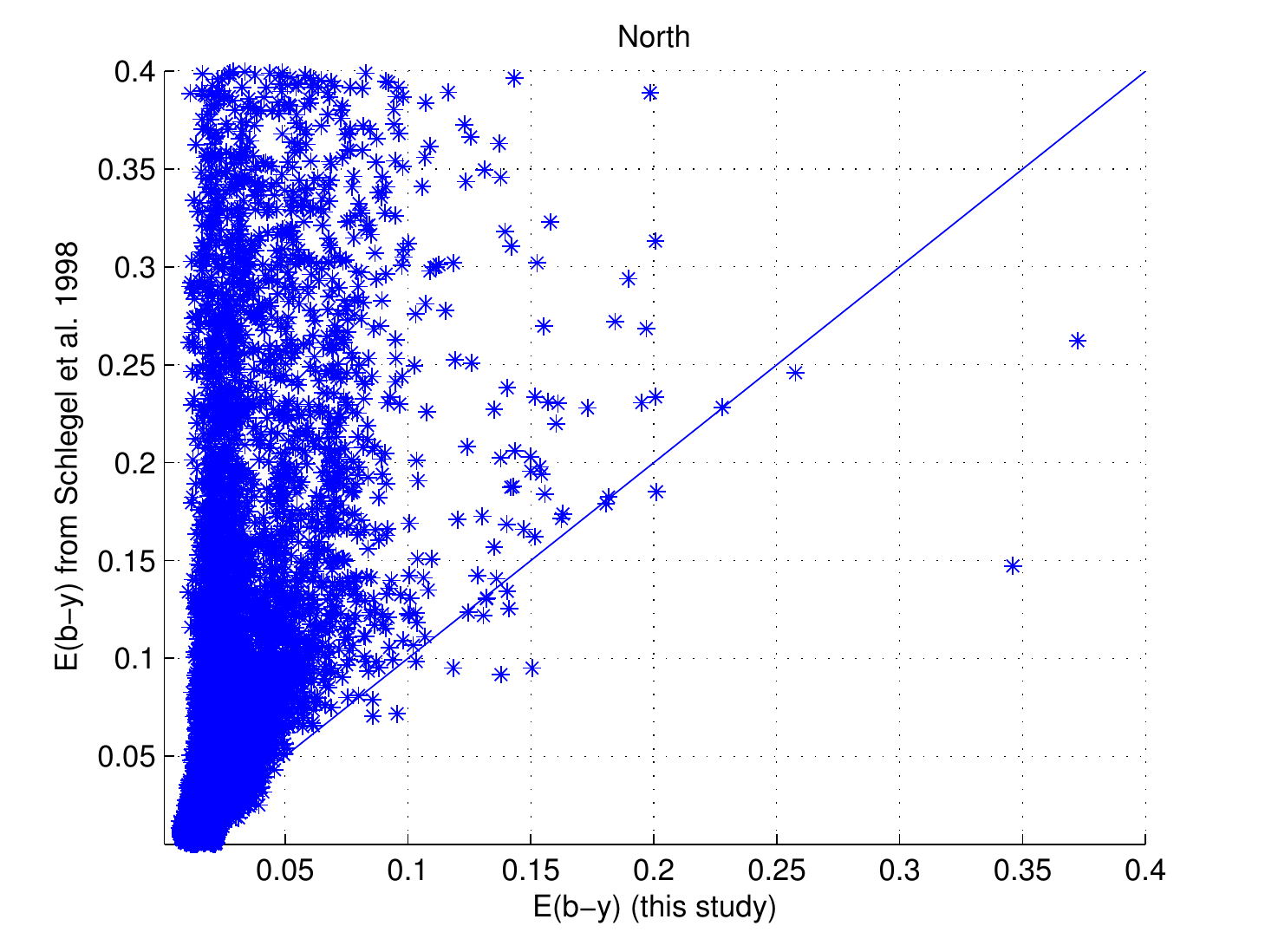}
  \caption{Comparison between Schlegel et al. extinction and extinction obtained in this study : north galactic pole in Lambert projection.
          }
     \label{schlegel1}
\end{figure}

\begin{figure}
\centering
\includegraphics[width=9cm]{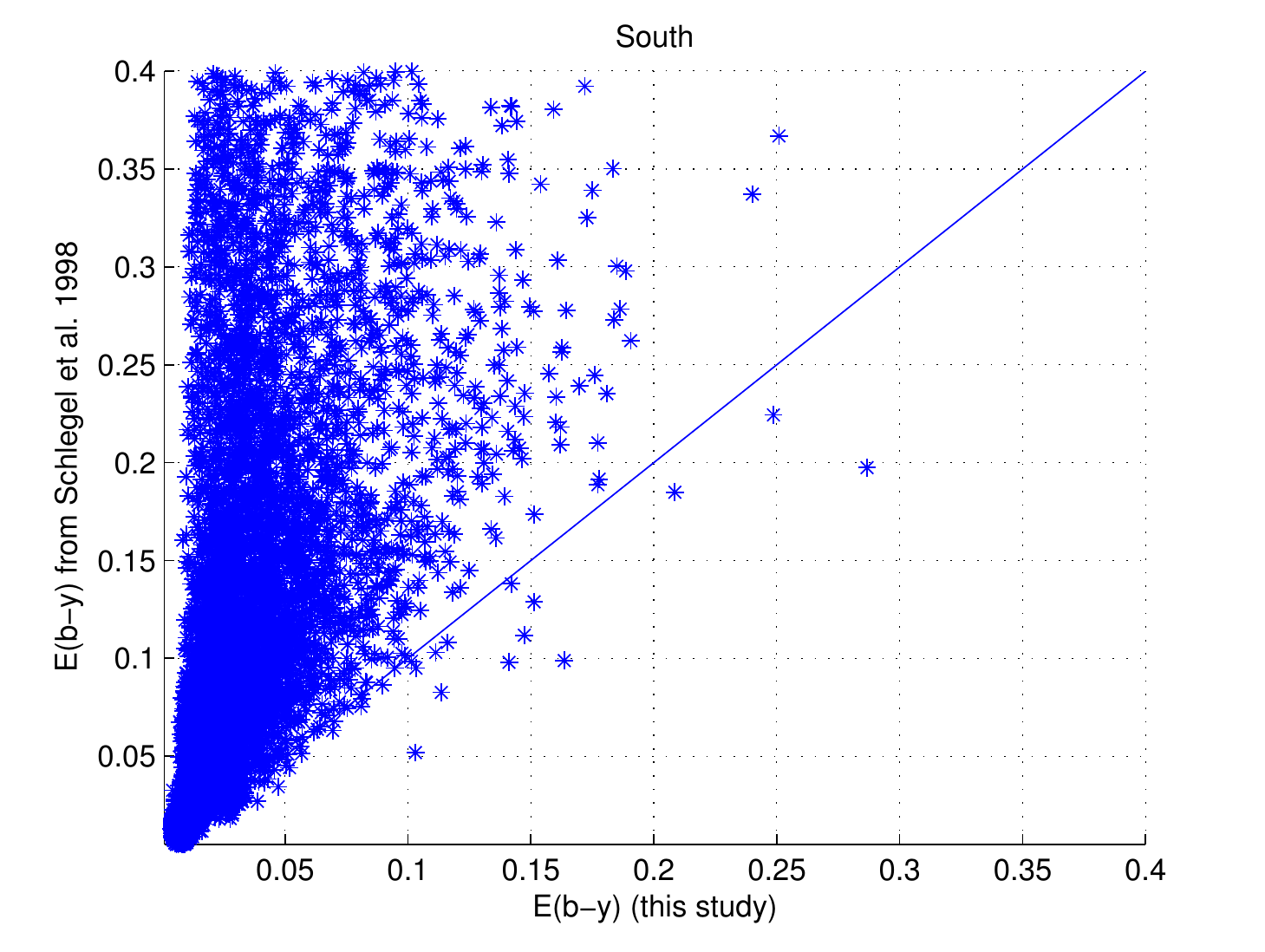}
  \caption{Same as Fig. \ref{schlegel1} for the south galactic pole.
          }
     \label{schlegel2}
\end{figure}

\begin{figure}
\centering
\includegraphics[width=9cm]{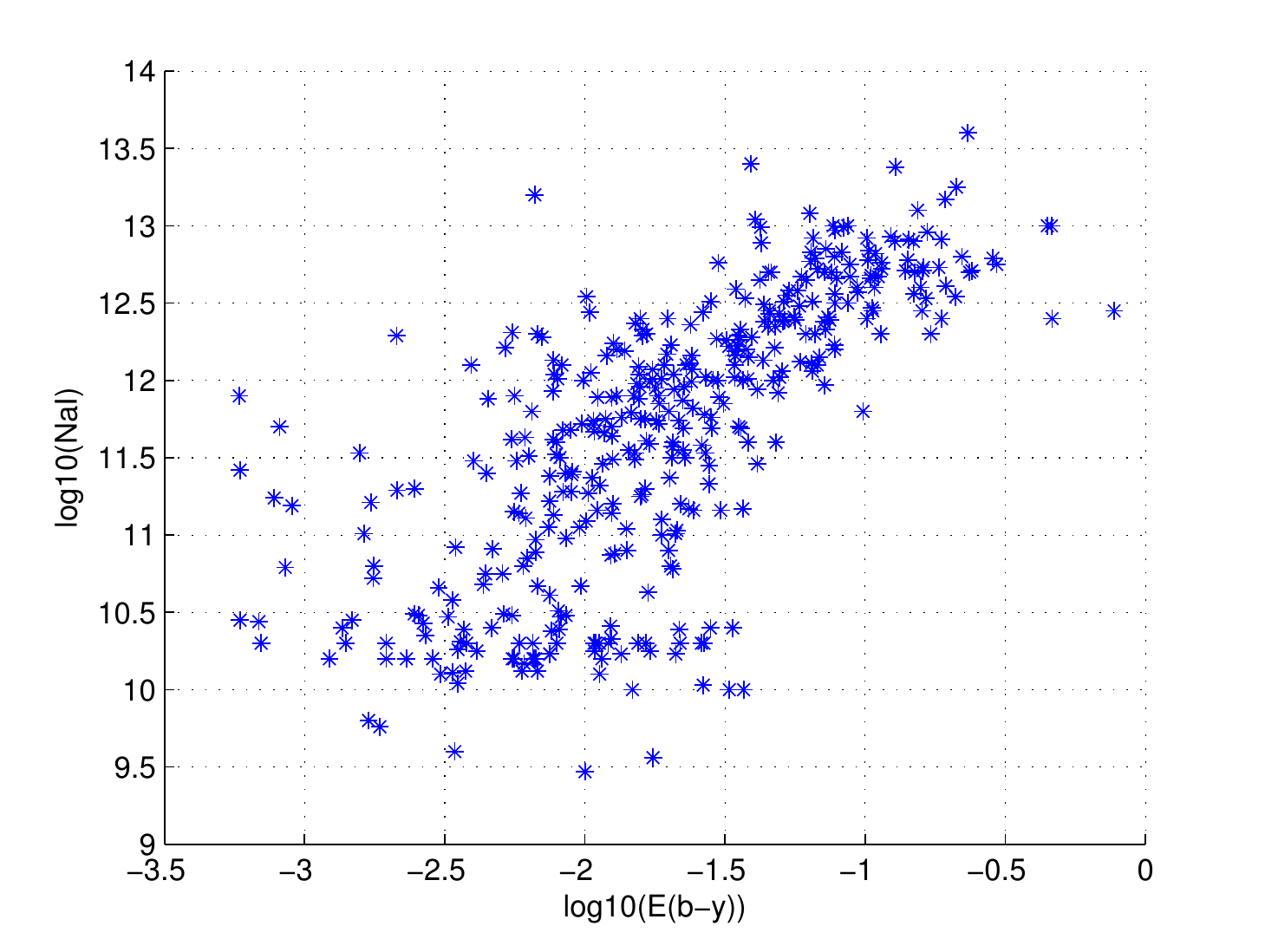}
  \caption{Sodium column density versus E(b-y) for common line of sight. }
     \label{fitextna}
\end{figure}


\section{Tomography of the opacity}

\subsection{The inverse method}
The determination of the 3-D scalar field from discrete
observational data is clearly under constrained because the
sample of stars is limited to a finite number of lines of sight
relatively spread-out. The problem can be regularized by
imposing that the true solution is smooth in some suitably
quantified sense (Twomey 1977; Craig \& Brown 1986). Thus one
reduces the number of free parameters, and the problem becomes
well-posed.

Here, we follow the approach of Tarantola \& Valette (1982a,b).
In this stochastic approach the prior information is
represented by probability measure for both data and model
parameters.  Thus true data $d$ and model $m$ are considered as
random vectors to take account of the errors in the data, in
the one hand, and of the "a priori" information on the model, in
the other hand. The inversion process consists in imposing that
the data are in fact linked to the model through a theoretical
mapping $g$ : $d = g(m)$. This is achieved by considering, as
posterior information on the model, the conditionnal
probability law for $m$ given that the a priori random vector
$d - g(m)$ is in fact null. In the present case of extinction,
$m$ is the opacity (or the log opacity) function over the
space, and $g$ is given by (eqn. 8). When both data and model
are assumed to be Gaussian random vectors, with centre and
covariance operator respectively denoted by $d_{prior}$, $C_d$,
and $m_{prior}$, $C_m$, and when $g$ is weakly non-linear,
Tarantola and Valette (1982b) showed that the posterior measure
for $m$ can be well approximated by the Gaussian measure, the
centre $\hat m$ of which makes minimum $E(m)$:
\begin{equation}
E(m) = \|C_m^{-1/2}(m - m_{prior})\|^2 + \|C_d^{-1/2} (g(m) -d_{prior})\|^2
\end{equation}
and the covariance operator of which is expressed as :
\begin{equation}
C = (C_m^{-1} + G_m^{*}C_d^{-1}G_m)^{-1}
\end{equation}
where  $G_m^{*}$ denotes the adjoint - with respect to the
usual scalar product of the model space and the data space - of
the derivative $G_m$ of $g$ at model m. The model $\hat m$
verifies the following equation (Tarantola and Valette, 1982b)
:
\begin{eqnarray}
\hat m - m_{prior} = C_m G_{\hat m}^{*}( C_d +
G_{\hat m} C_m G_{\hat m}^{*})^{-1} \nonumber \\
(d_{prior} - g(\hat m) 
+ G_{\hat m}(\hat m - m_{prior}))
\end{eqnarray}
which can be solved iteratively (Tarantola and Valette, 1982b) following a fixed point algorithm :
\begin{eqnarray}
m_{k+1} - m_{prior} = C_m G_k^{*}( C_d + G_k C_m G_k^{*})^{-1} \nonumber \\
(d_{prior} - g(m_k) 
+ G_k(m_k - m_{prior}))
\end{eqnarray}
where G$_k$ denotes the derivative of g at model m$_k$. It is
important to note that even though the model (i.e. the
log-opacity $\alpha$) is a function of the space, an iteration
only requires to solve a linear system with rank equal to the
number of data. From a practical point of view, the covariance
operator of the log-opacity is defined as $f \rightarrow C_m f$
with :
\begin{equation}
 C_m f (x) = \sigma(x) \int_V  \sigma(x')\, \psi(x,x') \,f(x')\, \d V(x')
\end{equation}
where $\psi(x,x')$ is the correlation of the log-opacity between point x and x', and where $\sigma^2(x)$
is its variance at point x. Hence, it follows that the algorithm (15) takes the following simple form :
\begin{equation}
\alpha_{k+1}(x) = \sigma(x) \sum_{i=1}^n \delta^i \int_0^{R_i} \rho(x_i)\,\sigma(x_i)\,\psi(x,x_i)\, \d r_i
\end{equation}
where $n$ is the number of data, and where the $n$-vector $\delta$ is solution of the following linear
system :
\begin{equation}
S_{ij}\, \delta^j = E_i + \int_0^{R_i} \rho_k (x_i)\,(\alpha_k(x_i) - 1)\, \d r_i
\end{equation}
with
\begin{equation}
S_{ij} = (C_d)_{ij} + \int_0^{R_i}\int_0^{R_j}\sigma(x_i)\,\rho_k(x_i)\,\sigma(x_j)\,
\rho_k(x_j)\,\psi(x_i,x_j)\,\d r_i \d r_j
\end{equation}

\subsection{Resolving power and averaging index}
  The concept of resolving kernel, introduced by Backus \& Gilbert (1970), is an effective tool
  to appreciate the efficiency of inversion. Taking that $d = g(p)$ into account, a first order expansion of
  $g$ around $m = \hat m$ yields
 \begin{equation}
 d_{prior} - g(\hat m) + G_{\hat m}(\hat m - m_{prior}) =
d_{prior} - d + G_{\hat m}(m - m_{prior})
\end{equation}
Substituting this expression into (eqn 14) gives :
\begin{equation}
\hat m - m_{prior} = R_{\hat m}(m - m_{prior})
- L_{\hat m}(d - d _{prior})
\end{equation}
where R$_{\hat m}$ is the resolution operator :
\begin{equation}
R_{\hat m} = C_m G_{\hat m}^{*}( C_d +
G_{\hat m} C_m G_{\hat m}^{*})^{-1} G_{\hat m}
\end{equation}
and
\begin{equation}
L_{\hat m} = C_m G_{\hat m}^{*}( C_d +
G_{\hat m} C_m G_{\hat m}^{*})^{-1} \mbox{  ,  } R_{\hat m} =
L_{\hat m} G_{\hat m}
\end{equation}
Expression (21) shows that, correct to first order, the model
$\hat m - m_{prior}$ is the sum of a term which corresponds to
the difference between the true model and the prior $m - m_{prior}$, filtered through the resolution
operator, and of a noise term coming from the discrepancy
between true and prior data. This expression can be rewritten
as:
\begin{equation}
(\hat m - m_{prior})(x) = \int_V h(x,x')\,(m - m_{prior})(x')\, \d V(x') + \mbox{ noise}
\end{equation}
where $h(x,x')$ is the kernel of the resolution operator, which is called the resolving, or the averaging kernel. Clearly, the value of $\hat m - m_{prior}$ at point $x$ is an average of the values of the
true model, weighted by the kernel, all around, and the resolution would be perfect if this kernel was
equal to the Dirac's distribution $\delta(x-x')$. Given a point $x$, $h(x,x')$ is generally a peaked function
of $x'$, centered around $x'=x$, with few negative lobes, and there are many ways to define its width (Rodgers,
2008, p. 54).

Here, we will focus on another important and useful concept, closely related to the averaging kernel.
Let ($\overline{m - m_{prior}}$) be the mean value of the true model within the width of the kernel around $x$.
Equation (24) reads :
   \begin{equation}
 (\hat m - m_{prior})(x) = I(x)\,(\overline{m - m_{prior}}) + \mbox{ noise}
   \end{equation}
 where :
   \begin{equation}
      I(x) := \int_V h(x,x')\, \d V(x')
   \end{equation}
 is called the {\it averaging index}. In those areas where the data provides poor information,
 $I(x)$ takes very low values, and hence, the resulting model remains close to the prior one,
 independantly of the value of the true model in the neighbourhood. In contrast, where the index
 is close to one, the resulting model corresponds to the mean value of the true model within the width
 of the averaging kernel. Thus, displaying the area where the index is low is a way to indicate the
 limit of the zone where the inference becomes poor.


\subsection{Prior information on the opacity distribution.}
In this method the model parameters may include
finite dimensional vectors as well as functions. For finite dimensional model spaces,
$C_m$ is simply a matrix. In the case in which we do not have any prior information on parameters,
the parameters must be a priori decorrelated with large variances.
In the case of functions, we must define the correlation kernel $\psi(x,x')$ of the covariance operator (16).
Simple kernels are Gaussian kernels
\begin{equation}
\psi (x,x')  =  \exp\left(- \frac{\| x - x'\|^2}{2\xi_{0}^2}\right)
 \end{equation}
or exponential kernel :
 \begin{equation}
\psi (x,x')  =  \exp\left(- \frac{\| x - x'\|}{\xi_{0}}\right)
 \end{equation}
where $\xi_0$ is a smoothing length that controls the spatial variability of the random function.\\
In this study, we first used a Gaussian kernel that yielded a
poor data fit. Since interstellar medium clearly shows
different characteristic lengths, we found more judicious to
use multiscale kernels (Serban and Jacobsen (2001)). Actually,
the introduction of two scales, the first one characterizing
the diffuse matter, and the second one, the compact clouds or
group of clouds, was sufficient to obtain a good adjustment of
data, in a second step. The covariance kernel that we have used
is the following:
  \begin{eqnarray}
  \label{corr}
\sigma(x)\,\sigma(x')\,\psi(x,x')  =  \sigma_{0}(x)\,\sigma_{0}(x')\,
        \exp\left(-\frac{\| x - x ' \|}{\xi_{0}}\right) + \nonumber\\
        \sigma_{1}(x)\,\sigma_{1}(x')\,
        \exp\left(-\frac{\| x -x'\|}{\xi_{1}}\right)
  \end{eqnarray}
The characteristic lengths $\xi_{0}$ and $\xi_{1}$ were taken
respectively equal to 15 and 5 pc. The levels of fluctuation
$\sigma_{0}$ and $\sigma_{1}$ were taken equal to 1.5 and 0.8
respectively. The choice of this double scale is important: it
corresponds to the two observed physical characteristics of the
main phases we are tracing: the warm diffuse gas, which is
widely distributed with low density, and the cold atomic phase
which is composed of smaller and denser clouds.

\subsection{Data adjustment.}

The extinction data we used are not complete, in the sense that
the more extinguished stars are not observed because of
magnitude selection. In the core of dense clouds, extinction
reaches 10 mag (Cambresy, 1997) with typical cloud sizes lower
than a few pc. Thus, we underestimate the opacity in the
darkest regions of the sky and the star sample used in this
study allows us to detect only diffuse clouds. Moreover, the
real density fluctuations occur at different scales and the
smoothing length allows us to detect only typical dust cloud
sizes larger than 5 pc which corresponds to the smallest
correlation length used in Eqn. \ref{corr}. The spatial
fluctuations smaller than
this smoothing length are not detectable.\\
Some extinction data are not compatible with our model (Fig
\ref{fit}), more particularly stars with large observed
extinction. The colors of these stars are probably affected by
structures smaller than the correlation length and cannot
be taken into consideration in our model.\\
For this reason, Figure \ref{fit} shows asymetric behaviour:
some large observed extinctions are not correctly adjusted by
our model and are underestimated. An other possibility is the
presence of outliers due to the fact that peculiar calibration
stars could not be detected during selection step.

\section{Results}

The result of the inversion procedure is a 101$^{3}$ data cube, corresponding to a 500x500x500 pc$^{3}$  cube in physical space, each individual 5x5x5 pc$^{3}$ pixel being attributed a volume opacity. In this section we present the main characteristics of this opacity distribution, along with the results of a similar inversion of 1670 interstellar neutral sodium (NaI) column densities measured towards stars within 300 parsecs. The sodium data are those presented in Welsh et al, (2009). At variance with the inversion shown in Welsh et al, performed for a unique kernel of 20 pc correlation length, the present sodium inversion has been made based on the same data but for exactly the same inversion parameters as for the opacities, i.e. two kernels with correlation lengths of 5 and 15 pc respectively. One may expect some small differences from the Welsh et al maps, especially at the smallest scale, since the present inversion allows for smaller clouds. The choice of identical kernels is motivated by  the need to eliminate one of the source of potential differences between sodium and dust opacity.\\
This parallel analysis has two main reasons. First, both neutral sodium and opacity probe dense interstellar clouds and it is interesting to compare the distributions from the point of view of the ISM itself. The second reason, more important at this stage of our inversion studies, is technical. It is very informative to compare two inversions based on two completely independent data sets, because  it provides some tests of the inversion, especially about the choice of the kernels in relation with the target distribution and its statistical significance. As we discussed earlier, the set of target stars for the opacity study is characterised by a non homogeneous distribution, with some regions particularly well covered, such as the Southern hemisphere region at l$<20^{\circ}$ and l$>200^{\circ}$. Also, the density of stars decreases strongly beyond a distance varying between 100 and
200 pc depending on the directions, due to selection effects linked to very strong extinction. The set of sodium target stars on the other hand is more regularly distributed with distance, but is a factor of four smaller.  It is interesting to see the effect of such different grids on the large scale distribution of the inverted quantity.\\

\subsection{Integrated opacities}

\subsubsection{Comparison with Schlegel et al. full-sky dust maps}

An interesting comparison can be done using the reddening
maps of Schlegel et al. (\cite{shlegel}). These two maps (for the Northern and the Southern galactic hemispheres) give the total reddening
$E(B-V)$ deduced from dust emission measurements and a temperature correction.
Using our extinction cloud model, we are able to estimate, at
high galactic latitude, the total expected extinction,
integrating Eqn. \ref{equa_fond2} from 0 to infinity (in
practice, the integration is stopped at 250 pc). In order to
compare directly
the results of our model with those of Shlegel, we use the same grid representation and interpolate the integrated extinction on a similar sampling. \\
Figures \ref{schlegel1} and \ref{schlegel2} show the
comparisons separately for the northern and southern galactic
hemisphere. This shows clearly that the extinction results are
compatible: the extinction structures recovered from stars are
underestimated according to the fact that cloud cores are not
tracked by our sample.


\begin{figure}
\centering
\includegraphics[width=9cm,height=10cm]{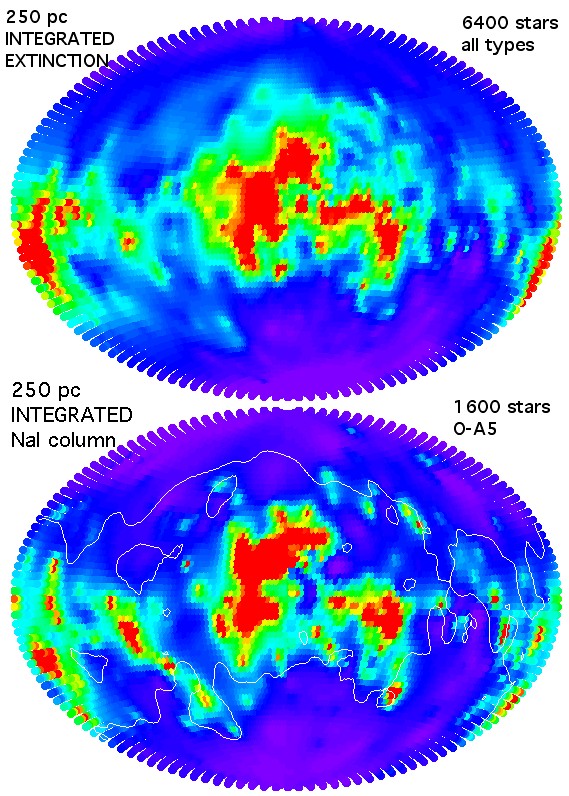}
\caption{Integrated opacity (resp. NaI column density) along 250 parsec long lines-of-sight based on the inverted cubes, here represented in Aitoff coordinates.}
\label{nadustaitoff}
\end{figure}

\subsubsection{Comparison with integrated NaI columns}
A first global comparison between the two inversions can be done by integrating back the cube values, i.e. the volume opacities and NaI densities, from the center to the surface of the cubes. Starting with the inverted cubes we have computed, for a regular grid of directions and for all the sky, the total opacity and the total NaI column integrated from the Sun up to a distance of 250 pc. The results are displayed in Fig \ref{nadustaitoff} for a grid of directions evenly spaced by 2.5 $\deg$.\\
A first look at those figures reveals immediately the local galactic disk "warp", clearly visible here in both the dust and gas distributions, with a general trend of concentration above the plane towards galactic longitudes 325 and 75 degrees, and below the plane elsewhere. This corresponds to the well-known Gould belt distribution of nearby O-B stars and gas. Such a clearly reconstructed "warp" in the two maps is a very encouraging result since the "a priori" solutions in both cases are plane-parallel and symmetric with respect to the plane.
The similarities between the two integrated quantities are also obvious when considering the large scales. This demonstrates that the stellar grids are dense enough to reveal the main concentrations of dust and gas.  A closer inspection reveals some differences at the smaller angular scales, especially in the very dense regions. Such differences will be more clearly visible in the 2D cuts through the cubes presented below, and may correspond to (i) inaccuracies of the inversion due to a too coarse stellar grid, (ii) actual differences between gas and dust, (iii) inaccurate data. The detailed study of all these  discrepancies is a very long task which is progressing along with the inversion of increasing amounts of targets. \\
A clearly revealed feature is the marked difference between the integrated opacity at the North and South poles. This is to relate to the North-South asymmetry already found by Schlegel et al (1998) in their reconstructed dust extinction maps based on COBE/DIRBE and IRAS data. That there is significant dust and a high dust/gas ratio in the North has also been inferred from extinction and HI data by several works (Knude and H{\o}g, 1999). \\
Interestingly, the North-South asymmetry is not obviously present in the case of the sodium columns, and the comparison between the two maps suggests that the dust/gas ratio is different in the two polar areas within 250 parsecs. Such a difference, which should have some consequences in the context of dust emission models, certainly deserves further studies.

\begin{figure*}
\centering
\includegraphics[width=14cm,height=7cm]{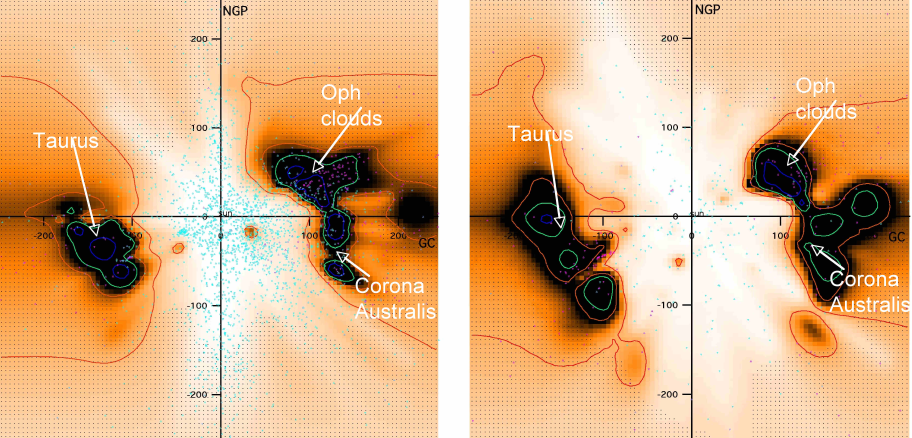}
  \caption{Volume opacity (left) and NaI volume density (right) within the meridian plane. Added are several iso-contours, as well as the locations (triangles) of the target stars close to the plane (see text). The marker colour represents the integrated opacity (resp. column) towards the star. Small dots are plotted in areas where there are no constraints on the inversion, and the volume opacity (resp. density is equal to the a priori value.}
     \label{merextna}
\end{figure*}

\begin{figure*}
\centering
\includegraphics[width=14cm,height=7cm]{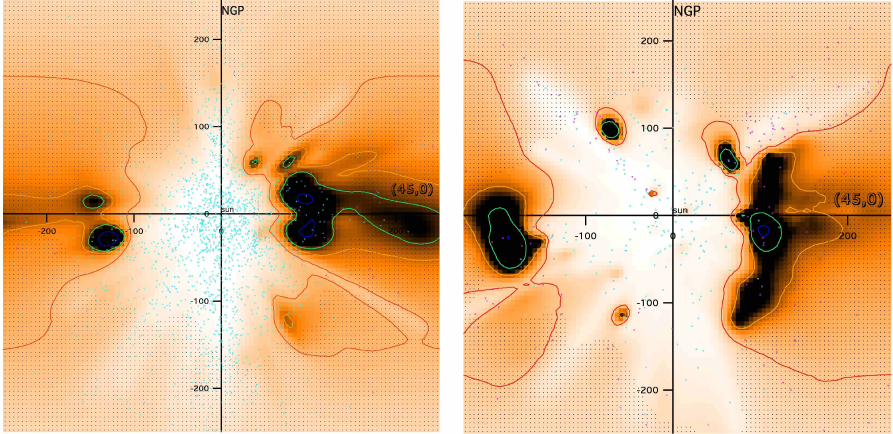}
  \caption{Same as figure \ref{merextna}, within a vertical plane at 45$\deg$ from the meridian. }
     \label{vert45extna}
\end{figure*}

\begin{figure*}
\centering
\includegraphics[width=14cm,height=7cm]{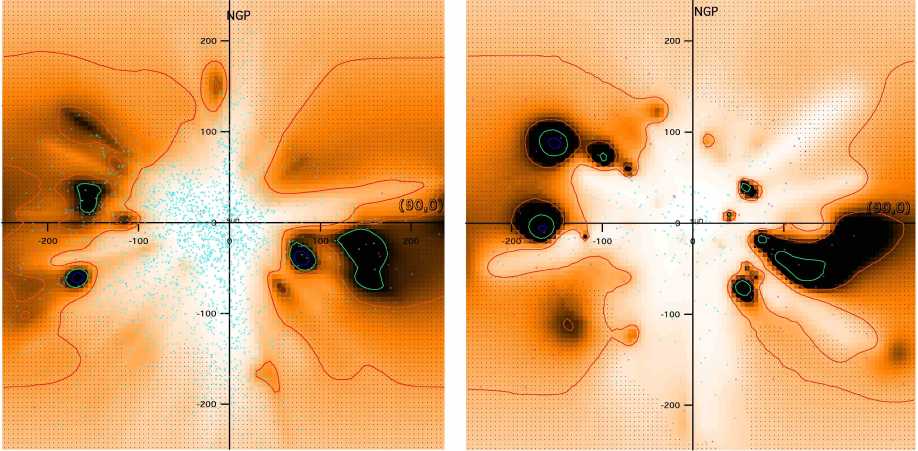}
\caption{Same as figure \ref{merextna}, within the rotation plane.}
\label{rotextna}
\end{figure*}

\begin{figure*}
\centering
\includegraphics[width=14cm,height=7cm]{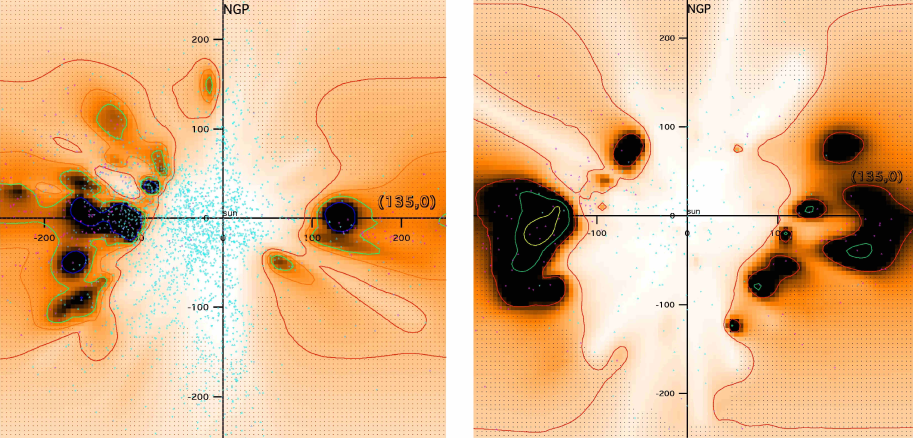}
  \caption{Same as figure \ref{merextna}, within a vertical plane at 45$\deg$ from the rotation plane }
     \label{vert135extna}
\end{figure*}

\subsection{Cuts in the 3D opacity and density cubes, identification of structures}

In this section we show the volume opacity and NaI density  within several planes, i.e. different cuts in the 3D inverted cubes. For each map is shown the continuous value of the volume opacity (resp. density) with some iso- volume opacity (iso-volume density) contours drawn to help visualise the dense areas and the voids. Also shown are the stars which are close to the plane and have contributed to the adjustment. The criterion for keeping the stars is identical to the one presented in Sfeir et al (1999). The colour scale for the stars corresponds to the measured integrated opacity (resp. column density) between the Sun and the star. This  allows to figure out how many stars have constrained the inversion and which stars have most influenced the local opacity (density). \\
Also shown in the maps are some markers of the reliability of the inversion process, which is directly linked to the number of targets and their spatial distributions: areas with averaging index lower than 0.5 are marked by black dots. In almost all maps, the inversion is well constrained within 100-150 parsecs, while beyond this distances there are large differences according to the longitude or latitude. The most constrained plane for neutral sodium is the Gould belt plane, since there are numerous nearby early-type stars along the belt, which are ideal targets. High-latitude areas are not well sampled, at the exception of the poles themselves or well known "windows" of low columns.

\subsubsection{Vertical planes}
Figures \ref{merextna} to \ref{vert135extna} display the inverted volume opacity and Na density in four vertical planes containing the Sun at 45$\deg$ from each other, starting with the meridian plane, i.e. the vertical plane which contains the galactic centre and anti-centre directions.
Such vertical planes are the most appropriate cuts in the 3D cubes for revealing the limits of the inversion process linked to the data sets, since the space density of the targets decreases strongly with the distance to the plane. Indeed, the influence of the "a priori distribution" can be easily discerned in those maps in all unconstrained areas, under the form of a smooth decrease of the opacity (density) as a function of the distance to the plane and horizontal iso-densities. This is a visible criterion of lack of constraints.

On the other hand, the juxtaposition of the same maps drawn from opacity and neutral sodium also clearly reveals the limitations of the 3D reconstruction. In some locations there are some differences between the centroids of the dense areas, or the absence of a feature in one of the two maps. Those shortcomings are related to the target scarcity in some areas, or may also be related to parallax uncertainties.\\
Despite those limitations, the Local Cavity is clearly defined in each of those vertical planes, and as noted in our previous works, is significantly inclined with respect to the plane. The global inclination reaches maxima in the 0-180$\deg$ meridian plane and in the  315-135$\deg$ vertical plane, and the maps suggest a complicated structure with a number of low-density chimneys linking the cavity to the halo. \\
Some nearby cavities are revealed by those vertical maps, almost in all cases in both extinction and gas maps when there are constraints in those areas, but frequently the dust or gas distributions around the cavities differ slightly or more significantly, due to the target limitations.
Work is in progress to identify these cavities in radio and X-ray maps and will be the subject of a forthcoming paper (Raimond et al, 2010).
We have identified a few structures in the meridian plane in Fig \ref{merextna}. The distances to Taurus and $\rho$Oph clouds will be discussed in the next section. The nearby Corona Australis concentration of clouds at -20$\deg$ is present  but there are some substantial differences in the structure derived from extinction and the sodium inversion. These differences may be due to the scarcity of targets stars in this area, and the small size of the CrA molecular region. The distance to the densest area in this direction is 135 pc from extinction, and about 165pc on the sodium map.
The former distance corresponds to previous estimates, and the latter to the estimate by Knude and H\"{o}g (1998). More data are required in this area.

\begin{figure*}
\centering
\includegraphics[width=14cm]{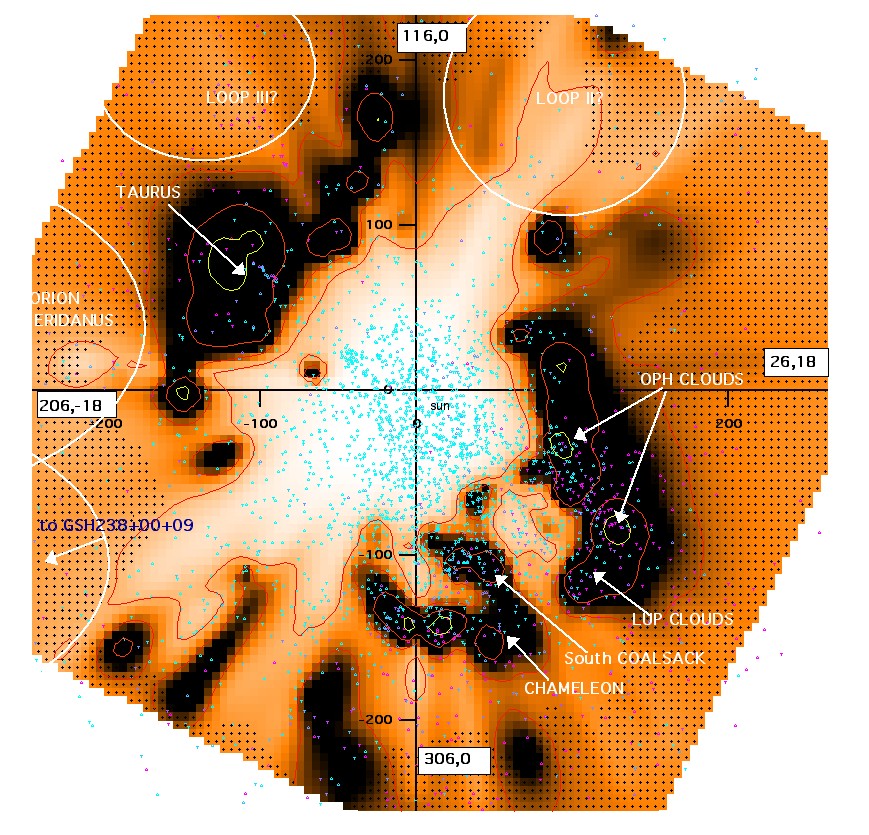}
\caption{Volume opacity within the Gould Belt plane (orientation taken from Perrot and Grenier, 2003}
\label{GBext}
\end{figure*}

\subsubsection{The Gould belt/Lindblad ring  plane}

\begin{figure*}
\centering
\includegraphics[width=14cm]{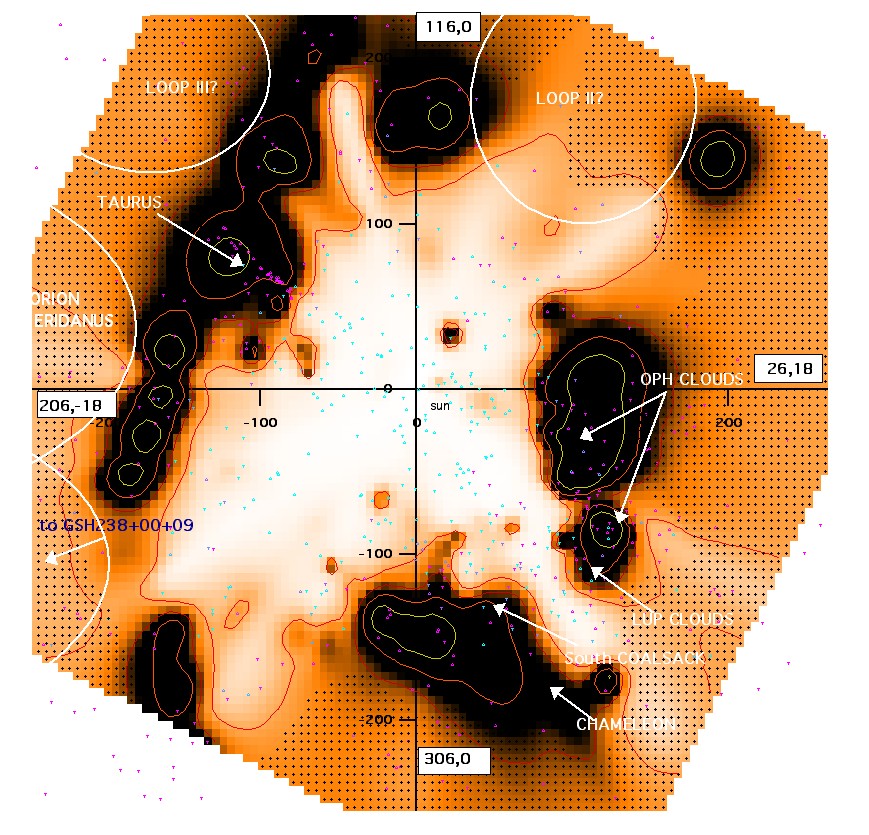}
\caption{Same as figure \ref{GBext}, for the neutral sodium volume density.}
\label{GBna}
\end{figure*}

In opposition to the vertical planes, the galactic plane and the Gould belt/Lindblad ring (GB)  plane (see P\"{o}ppel (1997) for a recent review on this structure) are among the most favourable regions for the inversion from the point of view of target distribution. For the extinction stellar dataset, the galactic plane is the most appropriate, but the Gould belt plane, being only slightly inclined, also contains a high density of targets. For the sodium absorption data set, it is the most appropriate, since it contains a large amount of bright, early-type stars that constitute ideal targets.
On the other hand, the GB plane is highly structured since it contains most of the dark clouds, star-forming regions and HII regions. Although this complexity makes the inversion more difficult, it is interesting to attempt a mapping within this plane.
We have thus chosen to display the inversion results within this particular plane, and the corresponding cuts in the two inverted cubes are displayed in Figures \ref{GBext} and \ref{GBna}.\\

The Gould belt/Lindblad ring structure has been the subjects of extensive studies, and we have chosen here for the definition of the plane the coordinates calculated by Perrot and Grenier (2003), based on the distribution and kinematics of OB associations  and HI and H$_{2}$ clouds.
It is generally admitted that the belt, whose origin is still controversial, is about 600 pc wide and that the Sun lies within about 150 pc from its closest part, the Scorpius-Ophuchus complex of OB associations and dense clouds.
The opacity and gas maps both reveal this complex area, and locates the East and West Ophiuchus clouds and the Lupus clouds from 100 to 165pc, in agreement with the recent work of Snow et al (2008) from absorption data (122 $\pm$8 pc for the densest part of $\rho$Oph and 133-150 pc for its western part), and Lombardi et al (2008) from reddening data (119 $\pm$6 pc for the densest part of $\rho$Oph and 155 $\pm$8 pc  for Lupus). The advantage of the present map compared to these two specific studies is that it provides a global view of all this region and the continuity between these regions. Similarly the map reveals the structure of the  Coalsack (its southern part only is within this plane) and Chameleon dark clouds at distances similar to the previous results of Knude and H\"{o}g (1998). These authors found between 100 and 150 pc for South Coalsack, which corresponds well to the elongated dense area in our map, and two separate dense regions  for Chameleon, the densest at about 150 pc and the second much closer, starting at about 80 pc.  This again corresponds well to the  double structure we find in this direction. Opposite to this dense region, is the Taurus dark cloud complex, at about the same distance (150pc), in agreement with previous work on the young objects and molecular clouds in this area (\cite{bertout}). The star distribution allows to detect beyond Taurus the closest part of the giant Orion-Eridanus superbubble, along with some nearby condensations at its periphery. The Local cavity is opening to another giant cavity, the GS238+00+09 superbubble (Heiles, 1998), adjacent to Orion-Eridanus.
Interestingly, the Local Cavity has two extensions at l=85$\deg$ and l=240-250 $\deg$ in this plane which are nearly perpendicular to the axis defined by the densest regions Taurus and Sco-Oph. The longest extension co\"{i}ncides with the well-know Canis Major extended void.  \\

Again, the comparison between the two distributions shows a rather good agreement about the global structure, which shows that the two datasets are good enough for such a large-scale determination. On the other hand, there are differences in the detailed structure. Work is in progress to identify the molecular complexes detected in radio data, and preliminary results confirm that, while some of the smallest isolated clouds are missed due to insufficient sky coverage, the main concentrations are retrieved. Further investigations are necessary in those cases, both the opacity and sodium star distributions are generally not sufficient to answer. The most important difference is at l=235$\deg$, at a distance of about 200 pc. Here the dense concentration detected in sodium does not seem to have a similarly strong counterpart in opacity. It is the contrary in the opposite direction, at about 120 pc. Is this due to a strongly variable dust to gas ratio, or to an artefact? \\

From those maps and the distribution of the quality index, it is clear that there is a lack of targets beyond the densest concentrations. E.g. it is not clear where the dust and gas densities drop beyond the Sco-Oph and Lupus clouds, or beyond Taurus. This is an important aspect, since the precise localization of the nearby cavities, potential X-ray emitters, would be a valuable advance.

\section{Conclusions}

We have built a carefully and conservatively selected  list of opacity measurements based on Str\"omgren photometry, restricted to nearby stars possessing a Hipparcos parallax and located within 300 pc. We have inverted this opacity database (6400 targets) together with the interstellar sodium absorption database (1650 targets) presented in Welsh et al (2009) by means of an inversion tool specifically adjusted to those data. We have compared the resulting 3D dust and gas distributions in different ways, including maps of integrated opacity and density as well as planar cuts within the 3D cubes. \\
The main conclusions of those inversions are:\\
-Integrals over 250 parsecs of the reconstructed opacities and sodium densities show very similar patterns over the sky, and reveal in a similar fashion the local warp of the plane, a warp link to the Gould belt-Lindblad ring global structure. \\
-The comparison with the integrated opacities from Schlegel et al. (1998)  shows that the present results are compatible with those obtained from dust emission maps.\\
-The extinction maps suggest a higher dust content at high galactic latitude in the Northern hemisphere compared to the South, while the gas maps do not. There is thus evidence for a higher dust to gas ratio in the North, in agreement with previous evidences (Knude and H\"{o}g, 1999)\\
-Both tracers clearly reveal the Local Cavity and its contours are similar, despite the totally independent, and indeed very different data sets.\\
-In general the large scale structures are in good agreement everywhere there are constraints from the data.\\
-At smaller scale there are some differences, that may be real but in most cases are likely due to the limited target coverage. This is clearly seen in the maps, especially far from the plane in the vertical cuts. Work is in progress on those discrepancies. \\
The results of those comparisons are encouraging with respect
to future work by showing that such inverse methods globally provide similar structures despite the total independence of the two data sets and the strong differences in the target distributions. They demonstrate that the production of
maps, with a precision of, say, 5 parsecs is not unrealistic.
The extensive parallax and opacities databases from the
forthcoming ESA mission GAIA and follow-up observations should
allow to obtain such a precision in the solar neighbourhood and to extend such maps to much larger distances.

\begin{acknowledgements}
We deeply thank our referee J. Knude for careful reading of our manuscript and constructive comments, as well as for his prompt review. 

This research has extensively used the Simbad database and on-line facility, operated by the CDS (Strasbourg, France).

\end{acknowledgements}

\end{document}